\documentclass{article}
\usepackage[]{psfig}

\newcommand{\be}{\begin{equation}}
\newcommand{\ee}{\end{equation}}
\newcommand{\ba}{\begin{eqnarray}}
\newcommand{\ea}{\end{eqnarray}}

\def\vr{{\check e}_r}
 \def\vteta{{\check e}_\theta}
 \def\vfi{{\check e}_\varphi}

\begin{document}
\title{Monopole Solutions in AdS Space}
\author{A.R.~Lugo\thanks{CONICET},
E.F.\ Moreno$^*$\,
and\,
F.A.\ Schaposnik\thanks{Associate CICBA}
\\
~\\
{\normalsize\it
Departamento de F\'\i sica, Universidad Nacional de La Plata}\\
{\normalsize\it
C.C. 67, 1900 La Plata, Argentina}
}
\date{\today}
\maketitle

\vspace{- 7 cm}
\hspace{8 cm}  {\small La Plata-Th 99/12}
\vspace{6.9 cm}

\begin{abstract}
We  find monopole  solutions for a spontaneously 
broken $SU(2)$-Higgs system coupled to gravity in asymptotically
anti-de Siter space.  We present new analytic and numerical results discussing,
in particular, how the gravitational instability of self-gravitating
monopoles depends on the value of the cosmological constant.
\end{abstract}

\bigskip



\section{Introduction}
Gravitating monopole solutions to gauge theories have attracted  many
investigations in the last 25 years \cite{BR}-\cite{VG}. In particular, the
existence  of self-gravitating monopoles in spontaneously
broken non-Abelian gauge theories, their properties, relation  with
black hole  solutions and their relevance in Cosmology have been
thoroughly discussed. Most of these investigations correspond
to asymptotically flat space-time but there have been  recently several
studies for the case in which the cosmological constant $\Lambda$
is non-vanishing
\cite{Wi}-\cite{LS}. In particular, we have discussed in \cite{LS}
the existence of gravitating  monopole solutions in the case in which 
space-time is asymptotically anti-de Sitter (AdS), which in our 
conventions corresponds to $\Lambda <0$.  Regular (monopole and dyon) and
singular (black hole) solutions have been found in this case and the
properties of the magnetically charged solutions  for vanishing Newton
constant $G$ were analysed.

It is the purpose of the present work to complete the 
investigation initiated in
\cite{LS} studying in detail the monopole solution in asymptotically AdS
space both for vanishing and finite $G$, analytically and numerically. The
plan of the paper is the following: we present  in Section II the model,
the spherically symmetric ansatz and the appropriate boundary conditions
leading to gravitating monopoles and dyons. Then, in Section III, we discuss,
analytically,
some relevant properties of the magnetically charged solution and then
describe in detail the numerical results   both
for $G=0$ and $G \ne 0$. We summarize and discuss our results in 
Section IV.

\section{The model}
We consider the action for $SU(2)$ Yang-Mills-Higgs theory coupled
to gravity in asymptotically anti-de-Sitter  space. The action
is defined as
\be
S = S_G + S_{YM} + S_H = \int d^Dx\sqrt{|G|} ( L_G + L_{YM} + L_H )
\label{1}
\ee
with
\be
L_G = \frac{1}{\alpha_0}  \left( \frac{1}{2} R - \Lambda \right)
\label{2}
\ee
\be
L_{YM} = -\frac{1}{4e{}^2}\; F_{\mu\nu}^a F^{a\,\mu\nu} \label{3}
\ee
\be
L_H = -\frac{1}{2}    D_\mu H^a
\; D^\mu H^a - V(H)
\label{4}
\ee
\be
 V(H) = \frac{\lambda}{4}\; ( H^a H^a - h_0{}^2 )^2
 \label{5}
 \ee
Here $F^a_{\mu \nu}$ ,
 ($a=1,2,3$)
  is the field strength,
\be
F^a_{\mu \nu} = \partial_\mu A_\nu - \partial_\nu A_\mu^a + 
\varepsilon^{abc}A_\mu^b A_\nu^c
\label{6}
\ee
and  the covariant derivative $D_\mu$ acting on the Higgs triplet $H^a$ 
is given by
\be
D_\mu H^a = \partial_\mu H^a + \varepsilon^{abc} A_\mu^b H^c
\label{7}
\ee
We have defined
\be
\alpha_0\equiv 8\pi G
\label{New}
\ee
where $G$ is the Newton constant,  $e$ the gauge coupling
and $\Lambda$ is the
cosmological constant (with our conventions
$\Lambda < 0$ corresponds, in the absence of matter, to anti-de Sitter
space).

The equations of motion that follow from (\ref{1}) are
\ba
E_{\mu\nu} + \Lambda\; G_{\mu\nu} &=&
\alpha_0 \; (T_{\mu\nu}^{YM} +  T_{\mu\nu}^H )\nonumber\\
 { D}_\rho D^\rho H^a &=& \frac{\delta V(H)}{\delta H^a}\nonumber\\
\frac{1}{e{}^2} D^\rho F^{a}_{\mu\rho} &=&
\varepsilon^{abc}\left( D_\mu H^b\right)    H^c
\ea
where $E_{\mu\nu}$ is the Einstein tensor and the matter energy-momentum tensor is
given by
\ba
T_{\mu\nu}^{YM} &=& \frac{1}{e{}^2}(- F_{\mu\rho}^a F_{a\;\nu}{}^\rho
+ \frac{1}{2}\; G_{\mu\nu}\; F_{\rho\sigma}^a F_a^{\rho\sigma} )\nonumber\\
T_{\mu\nu}^H &=&   \; D_\mu H^a\; D_\nu H^a + G_{\mu\nu}\; L_H
\label{T}
\ea

The most general static spherically symmetric form for the metric in $3$
spatial  dimensions together with the t'Hooft-Polyakov-Julia-Zee ansatz
for the gauge and Higgs fields in the usual vector notation reads
 \ba
 G &=& - \mu(x)\; A(x)^2\; d^2 t + \mu(x)^{-1}\; d^2 r + r^2\; d^2
\Omega_2\nonumber\\
\vec A &=& dt \; e\; h_0\; J(x)\; \vr - d\theta\; (1 - K(x) )\; \vfi
+ d\varphi \; (1 - K(x) )\; \sin\theta\; \vteta\nonumber\\
\vec H &=& h_0\; H(x)\; \vr
\label{ansatz}
\ea
where
we  have introduced the dimensionless coordinate $\; x\equiv e\; h_0\; r$
and $h_0$ sets the mass scale ($[h_0] = m^1$).

Using this ansatz, the equations of motion take the form
\ba
\left( x\; \mu (x)\right)' &=& 1  + 3\; \gamma_0\;  x^2
- \alpha_0 h_0{}^2
\left( \mu (x)\; V_1 + V_2 + \frac{x^2}{2}\;
\frac{J'(x)^2}{A(x)^2} +
\right.\nonumber\\
& & \left. \frac{J(x)^2 K(x)^2}{\mu(x) A(x)^2} \right)
 \label{mu}\\
x\; A'(x) &=& \alpha_0 h_0{}^2 \left( V_1 + \frac{J(x)^2 K(x)^2}{\mu(x)^2
A(x)^2} \right) A(x) \label{eqq}
\ea
\ba
\left( \mu (x) A(x) K'(x) \right)' &=& A(x)\; K(x) 
\left( \frac{K(x)^2
-1}{x^2} + H(x)^2- \right.\nonumber\\
&& \left. \frac{J(x)^2}{\mu (x) A(x)^2} 
\right)\label{K}\\
\left( x^2\mu(x) A(x) H'(x) \right)' &=& A(x)\; H(x) \left( 2\; K(x)^2
+\frac{\lambda}{e{}^2}\; x^2\; ( H(x)^2 - 1) \right)
\label{H}\\
\mu (x)\left( \frac{x^2 J'(x)}{A(x)} \right)' &=& \frac{2\; J(x)
K(x)^2}{A(x)}
\label{eqs}
\ea
where, for convenience, we have defined the dimensionless parameter
\be
\gamma_0 \equiv - \frac{\Lambda}{3 e{}^2 h_0{}^2}
\ee
  and
\ba
V_1 &=& K'(x)^2 + \frac{x^2}{2} H'(x)^2 \nonumber\\
V_2 &=& \frac{ (K(x)^2 -1)^2}{2\; x^2} +
\frac{\lambda}{4 e{}^2}\; x^2\; (H(x)^2 -1)^2
\ea

\subsection*{The boundary conditions}

Ansatz (\ref{ansatz}) will lead to  well behaved solutions for the matter
fields if, at $x=0$, one imposes
\begin{itemize}
\item ${H(x)}/{x}$ and  ${J(x)}/{x}$ are regular;
\item $1- K(x)$ and $K'(x)$ go to zero.
\item $\mu(x)  \to 1$

\end{itemize} 

On the other hand we want the system to go asymptotically to anti-de Sitter
space which corresponds to the solution of the Einstein equations with $\Lambda <0$
in absence of matter (see next Section); for
this to happen we must impose that the matter
energy-momentum tensor vanishes at spatial infinity.
From eq.(\ref{T}) one can see that the appropriate conditions
 for $x\rightarrow\infty$
are
\ba
A(x) &\rightarrow& 1  \nonumber\\
K(x) &\rightarrow& O(x^{-\alpha_1})\nonumber\\
H(x) &\rightarrow& H_\infty + O(x^{-1-\alpha_2})\nonumber\\
J(x) &\rightarrow& J_\infty + O(x^{-\alpha_3})
\label{boun}
\ea
with $\alpha_i > 0, ~i=1,2,3$.

Note that, being the equations for $A$ and $\mu$ first order, 
we impose just one condition for each one.

\section{The system in AdS space}

We shall first consider the case in which the Newton constant $G$ vanishes, 
so that
the gravitational
equations decouple from the matter and then  analyse the
full $G \ne 0$ problem. In the former case we 
have  already studied in \cite{LS}   the classical equations of
motion analytically, showing that monopoles could exist 
in asymptotically anti-de Sitter spaces and    
discussed  its main properties. We present in the next subsection the numerical
evidence that this solutions do exist, thus completing  the analysis 
in \cite{LS}. Then, we extend our study to the $G \ne 0$ case and 
again present both   analytical and numerical analysis
showing the existence of monopole solutions provided $G$ is
smaller than a critical value $G_c$.

\subsection*{The $G$ = 0 case}
Taking the  $\;\alpha_0 h_0{}^2 \rightarrow 0\;$ limit,
one easily finds for the metric
  the solution
\ba
A(x) &=& 1\nonumber\\
\mu(x) &=& 1 + \gamma_0 \; x^2 -\frac{a}{x}
\label{not}
\ea
which is nothing but the vacuum solution of the Einstein equations
with a cosmological constant (assumed negative), and corresponds
to a neutral Schwarzschild black hole in AdS space.
Concerning the integration constant $a$, it is
related to the mass of the black hole and
will be put to zero in what follows, in agreement with the
condition imposed on $\mu$ at $x=0$.
This metric, in turn, acts as a (AdS) background
with radius $r_0$,
\be
\; r_0 = \sqrt{-{3}/{\Lambda}}\;
\label{radius}
\ee
for the Yang-Mills-Higgs system.

For simplicity we   study  eqs.(\ref{eqs})
in the BPS limit which corresponds to
${\lambda}/{e{}^2} = 0$
with $h_0$ fixed.
\ba
( \mu(x) K'(x) )' &=& K(x) \left( \frac{K(x)^2 -1}{ x^2} + H(x)^2
 - \frac{J(x)^2}{\mu (x)} \right)\nonumber\\
( x^2\mu(x) H'(x) )' &=& 2\; H(x)\; K(x)^2 \nonumber\\
\mu(x)\; ( x^2 J'(x) )' &=& 2\; J(x)\; K(x)^2
\label{bps}
 \ea

The total amount of matter $M$ associated to the solution of (\ref{bps})
is defined as 
(see for example \cite{Gib})
\be
M = \int_{\Sigma_t}  d^3 x\; \sqrt{ g^{(3)}}\; T_{00}
\ee
where $\; g^{(3)} \;$ is the determinant of
the induced metric on surfaces $\Sigma_t$ of
constant time $t$ with normal vector
$\; e_0 = \mu(x)^{-\frac{1}{2}} \partial_t \;$ and
$\; T_{00}\equiv e_0^{\,\mu}\; e_0^{\,\nu}\; T_{\mu\nu} = {T_{tt}}/{\mu (x)}$
is the local energy density as seen by an observer moving on the flux
lines of $\partial_t$.
For the spherically symmetric configuration we are considering, it takes the
form
\be
M = \frac{4\pi h_0}{e}\; \int_0^\infty\; dx\;
\frac{x^2}{( 1+\gamma_0\; x^2)^\frac{3}{2} }\;
\frac{T_{tt}}{e{}^2\;h_0{}^4}
\ee
We quote for completeness
the explicit expressions for $T_{tt} = T_{tt}^{(YM)} + T_{tt}^{(H)} $
\ba
\frac{T_{tt}^{(YM)}}{e{}^2\;h_0{}^4} &=&
\frac{\mu(x)}{2}\; J'(x)^2 + \frac{J(x)^2\;K(x)^2}{x^2} +
\frac{\mu(x)^2\; K'(x)^2}{x^2} + \frac{\mu(x)}{2\;x^4}\; ( K(x)^2-1)^2\nonumber
\\
\frac{T_{tt}^{(H)}}{e{}^2\;h_0{}^4} &=&
\frac{\mu(x)^2}{2}\; H'(x)^2 + \frac{\mu(x)}{x^2}\; H(x)^2\; K(x)^2
\ea
It is not difficult to see from these expression
that the boundary conditions imposed through eqs.(\ref{boun})
are precisely those required for finiteness of $M$. 

As stated above, analytical arguments showing 
the possibility of monopole solutions were presented in \cite{LS}. 
To begin with, 
let us note that is possible to perform a power series expansion for   
large $x$ 
and calculate the coefficients in the expansion recursively.  
One can consistently propose an expansion  of the form
\begin{eqnarray}
K(x) &=& \frac{K_{\nu + 1}}{x^{\nu +1}}\; 
\left( 1 - \frac{k_\nu}{x^2} + \dots\right)\nonumber\\
H(x) &=& \frac{3\, H_3 }{x^3}\; \sum_{k=0}^\nu \; \frac{(-)^k}{2k+3}\,
\frac{1}{x^{2k}} + H_\infty \; 
\left( 1 + \frac{h_\nu}{x^{2\nu + 4}} + \dots \right)\nonumber\\
J(x) &=& \frac{J_1}{x} + J_\infty \; \left( 1 + \frac{j_\nu}{x^{2\nu + 4}}
 + \dots \right) 
 \label{exp}
\end{eqnarray}
and, after  insertion in eqs.(\ref{bps}), one can determine the
coefficients $k_\nu$, $h_\nu$ and
$j_\nu$ recursively. For positive integer $\nu$ one has
\begin{eqnarray}
k_\nu &=& \frac{\nu^2 + 3\, \nu + 3 + J_\infty{}^2}{2\, (2\nu + 3)}\nonumber\\
h_\nu &=& \frac{K_{\nu + 1}{}^2}{(\nu +2 ) \, (2\nu + 1)}\nonumber\\
j_\nu &=& \frac{K_{\nu + 1}{}^2}{(\nu +2 ) \, (2\nu + 3)}
\end{eqnarray}
Similar expressions can be obtained for $\nu$ a positive semi-integer. 

When such an expansion are assumed, non-trivial solutions exist if and only
if 
\be
H_\infty^2 = \nu(\nu+1) \gamma_0\; ,  \;\;\;  \nu=\frac{1}{2}, \, 1, \, \frac{3}{2}, \,  2, 
\, \ldots
\label{mmm}
\ee

If one relaxes a  power behavior like in (\ref{exp}), then one again
gets a relation like (\ref{mmm})
but
 with $\nu$ a real number, the leading exponent in the asymptotic expansion
 of $K(x)$. It is important to note that, having AdS space a natural 
 scale $r_0$, the system trades the in principle 
 arbitrary $h_0$ dimensionfull parameter for the AdS radius $r_0 = 
 \sqrt{-3/\Lambda}$ which now sets the scale, 
 \begin{equation} 
 |\vec H(\infty)|^2 = 
 \nu(\nu+1)\gamma_0h_0^2 = \nu(\nu+1)(er_o)^{-1}
 \end{equation}  
 Then, using (\ref{mmm})
 for $\nu$ real and a given $\Lambda$ is equivalent to consider
 an integer $\nu$  (for example  with $\nu=1$) 
 provided $r_0$ (i.e. $\Lambda$) is changed accordingly.

To obtain a detailed profile of the monopole solution, we solved
numerically the differential equations. For simplicity we considered
the $J=0$ case corresponding to a purely magnetic solution. The equations
of motion  read  

\ba
( \mu(x) K'(x) )' &=& K(x) \left( \frac{K(x)^2 -1}{ x^2} + H(x)^2
  \right)\label{Kr}\\
( x^2\mu(x) H'(x) )' &=& 2\; H(x)\; K(x)^2\label{Hr}\\
\mu(x) &=& 1 + \gamma_0 x^2 \label{mus}
\ea
We
employed a relaxation method for boundary value problems
\cite{NR}.  Such method determines the solution by starting with
an initial guess and improving it iteratively.  The natural
initial guess was the exact Prassad-Sommerfield solution \cite{BPS}
(which corresponds to $\gamma_0 = 0$).  
We have found regular monopole solutions for any value of the cosmological
constant $\Lambda$. We present in Figure 1 the solution profile for different
values of $\Lambda$. A distinctive feature of solutions for $\Lambda \ne 0$ 
compared with the flat-space Prasad-Sommerfield solution concerns the
asymptotic behavior of the fields. Indeed, when $\Lambda \ne 0$, the Higgs
field approaches its v.e.v. faster than in the Prasad-Sommerfield (PS)
case,
\be
H(x) \sim H_\infty + \frac{C_{\Lambda \ne 0}}{x^3} \, , \;\;\; x \gg 1
\label{C}
\ee
\be
~ \!\!\!\!H^{PS}(x) \sim H_\infty - \frac{1}{x} \, , \;\;\; x \gg 1
\label{Ci}
\ee
 As a result of this change of the asymptotic
 behavior, one can see that the    
 radius $R_c$  of the monopole core decreases. Indeed,
 as can bee seen in Fig.1,  
 when $\gamma_0$ (the cosmological constant) grows, $R_c$ becomes smaller 
 as the magnetic field concentrates near the origin.
 
We have also computed numerically the monopole mass which can be written as
 \be
M = \frac{4\pi}{e^2} \frac{1}{r_0} f_{{\gamma_0}}(\lambda/e^2)
\label{M}
\ee
where, extending the usual flat space notation, we have
introduced the dimensionless function $f_{{\gamma_0}}(\lambda/e^2)$. 
In the present case,  the cosmological constant provides
a natural scale and this has been exploited in (\ref{M}). This formula
can be written in units of the mass scale $h_0$ as
\be
\frac{M}{h_0} = \frac{4\pi}{e} E
\label{Mh}
\ee
where $E$ is a dimensionless function of $\gamma_0$,
\be
E = \sqrt{\gamma_0} f_{{\gamma_0}}(\lambda/e^2)
\label{E}
\ee
We present in Figure 2 a plot for $E$ as a function of $\gamma_0$ where  
one can see that 
$\lim_{{\gamma_0} \to 0} \sqrt{\gamma_0} f_{{\gamma_0}}(0) =
 1$ 
which is the correct result for Prasad-Sommerfield monopoles 
in flat space.

\subsection*{The $G \ne 0$ case}

In order to study the asymptotic behavior of the solutions to eqs.(\ref{eqq})
we have taken as independent metric functions $A(x)$ and $\tilde \mu(x) = 
A(x) \mu(x)$. Moreover, identifying $h_0  =(er_0)^{-1}$ ($\gamma_0 = 1)$,
the equations to study become, in the BPS limit, 

\begin{eqnarray}
A(x)\; \left( x\; \tilde \mu (x)\right)' &=& (1  + 3\;  x^2 ) \; A(x)^2
- \alpha_0 h_0{}^2\left( A(x)^2\; V_2 + \frac{x^2}{2}\; J'(x)^2\right)
\nonumber\\
x\;\tilde\mu (x)^2\; A'(x) &=& \alpha_0 h_0{}^2 \left( \tilde \mu(x) ^2 \; 
V_1 + J(x)^2 K(x)^2\right)
A(x)\nonumber\\
\left(\tilde  \mu (x) \; K'(x) \right)' &=& K(x) \left(  
 A(x)\frac{K(x)^2
-1}{x^2} +   A(x)\; H(x)^2- \frac{J(x)^2}{\tilde \mu} \right)\nonumber\\
\left( x^2\tilde \mu(x)\; H'(x) \right)' &=& 2\; A(x)\; H(x)\; K(x)^2
\nonumber\\
\tilde\mu (x)\; A(x) \left( x^2 J'(x) \right)' &=& x^2\; 
\tilde\mu(x)\; A'(x)\; J'(x) +
2\; A(x)^2\; K(x)^2\;J(x)
\end{eqnarray}
where, in the first one,  we have combined eqs.(\ref{mu}) and (\ref{eqq}).  

We consider for simplicity the purely magnetic case, $J=0$, and
propose a power series expansion of the form
\begin{eqnarray}
\tilde \mu (x) &=& x^2 + \sum_{m=0}^\infty\; \frac{\tilde\mu_m}{x^m}   
\;\;\; , \;\;\;\;\tilde \mu_0\equiv 1\nonumber\\
A(x) &=& \sum_{m=0}^\infty\; \frac{A_m}{x^m}   \;\;\; , \;\;\;\; A_0\equiv 1
\nonumber\\
K(x) &=& \left\{ \begin{array}{l}\sum_{m=0}^\infty\; \frac{K_m}{x^m}\;\;\; 
,\;\;\;\; K_0\equiv 0\\
\frac{1}{\sqrt x}\; \sum_{m=0}^\infty\; \frac{K_m}{x^m}\end{array}\right. 
\nonumber\\
H(x) &=& \sum_{m=0}^\infty\; \frac{H_m}{x^m}\;\;\;\; , \;\;\; H_0\equiv 
H_\infty \;\; , H_1\equiv 0
\end{eqnarray}
Again, coefficients can be determined recursively. The leading coefficients
in the expansions
  for $K$ and $H$ coincide with those already presented for $\alpha_0=0$
  (eq.(\ref{exp})).
We then just 
quote the corresponding ones  for the metric functions (for $\nu$ a
positive integer)
\begin{eqnarray}
\tilde \mu (x) &=&  1 +  x^2  + \frac{\tilde \mu_1}{x} + 
\frac{\alpha_0 h_0{}^2 }{2\; x^2} -
\frac{A_6}{x^4} + \dots\cr
A(x) &=& 1 + \frac{A_6}{ x^6} + \dots
\end{eqnarray}
where
$$
A_6 = - \alpha_0 h_0{}^2\; \left( \frac{3}{4} \, H_3{}^2 + \frac{2}{3}\, K_2{}^2\;\delta_{\nu ,1}      \right)
$$
We see that function $A$ is completely determined, to all orders, in terms of
$H$ and $K$ coefficients.  As an example, and from the numerical 
results described below, one finds for $\nu = 1$ and $\alpha_0 = 0.1$
that $\tilde \mu_1 = -0.24$,  $K_2 = 0.73$, 
$H_3 = -0.32$ and then $A_6 = -0.87$. 

Concerning the asymptotic expansion for $\tilde \mu$ corresponds to a
Reissner-N\"ord\-strom metric (with cosmological constant), 
with the free parameter
$-\tilde \mu_1$  related to the gravitatory mass and the coefficient
of the $1/x^2$ term, which arises for charged solutions, precisely corresponding
to the $Q_m=1$ magnetic solution we are considering. As seen from afar,  
and for an appropriate set of parameters, the metric can be identified with that of
a magnetically charged black hole as that described in \cite{LS},\cite{R}.
Concerning the expansion for $K$, it has one
free coefficient ($K_{\nu + 1}$),  while for the expansion for
$H$ two coefficients remain free ($H_\infty$
and $H_3$).

In order to get the detailed profile of the solutions, we have again to
solve numerically the equations of motion.
For simplicity, we have considered the BPS limit, $\lambda/e^2 = 0$.
Employing the same relaxation method as for the $G=0$ case
we have found a self-gravitating monopole solution
satisfying the boundary conditions previously discussed.  
Solutions   are similar to those corresponding
to  asymptotically flat space \cite{BFM1}-\cite{LW}.
In particular, we have found a maximum value for the
gravitational interaction strength $\alpha_0$ such that above  $\alpha_0^c$ 
the solution ceases to exist. This effect, already encountered
in asymptotically
flat space, can be understood noting that as $\alpha_0$ increases from $0$
to its critical value, the ratio ${\cal M} = mass/radius$ for
the monopole solution also 
increases
until it becomes gravitationally unstable. Now, as the cosmological constant
$|\Lambda|$ increases, the radius of the monopole decreases (the
behavior for $\alpha_0 \ne 0$ is analogous to that depicted in Fig.1
for $\alpha_0 = 0$)
while the mass of the monopole increases (the behavior
for $\alpha_0 \ne 0$ is
analogous to that in Fig.2) so that ${\cal M}$
is a monotonically growing function of $-\Lambda$ or, what is the same,
of $\gamma_0$. This explains why
the critical value $\alpha_0^c(\gamma_0)$  for $\gamma_0 >0$,   is 
smaller than the asymptotically flat one, 
$\alpha_0^c(\gamma_0)< \alpha_0^c(0)$: the critical value
${\cal M}_c$ at which the solution collapses is reached
before, in the $\alpha$ domain, for $\gamma_0  >0$ than for $\gamma_0 = 0$.  
As an example, for $\lambda = 0$  and $\gamma_0 = 1$ the
critical $\alpha_0$-value is $\alpha_0^c(1) = 1.374$ to be compared
with the asymptotically flat space value $\alpha_0^c(0) = 5.549$.
 
Concerning
the solution for the metric, as can be seen in Fig. 3,  
$\mu/(1 + \gamma_0 x^2)$ has a minimum which decreases as 
the strength of the gravitation interaction grows and tends to 
zero as $\alpha_0 \to \alpha_0^c$. 
A similar behavior can be seen to occur for
$A(x)$ which has also a minimum at the origin which tends to zero 
as $\alpha_0 \to \alpha_0^c$.  As in the case of asymptotically flat space,
$A$ develops a step-like behavior which becomes more and more sharp as 
$\alpha_0^c$ is approached. The position of the center of the step function
can be used to determine the corresponding value of the horizon, which
for AdS spaces results from the solution of a quartic algebraic equation.
\cite{R},\cite{LS}

\section{Discussion}
In this work we have studied
in detail the monopole and dyon solutions to Yang-Mills-Higgs
theory coupled to gravity for asymptotically anti-de Sitter space
presented in ref.\cite{LS}.  We have first considered the case in which the
Newton constant $\alpha_0$ vanishes so that the Einstein equations decouple leading to
a Schwarzschild black hole in AdS space. This metric acts as a background
for dyon solutions which were studied in detail making both an analytical and 
a numerical analysis. A distinctive feature of AdS solutions in this 
case is that the 
monopole radius is smaller than that corresponding to the $\Lambda = 0$
case. Apart from this property, qualitatively, the Higgs field and magnetic
field behavior is very similar to that corresponding 
to the 't Hooft-Polyakov solution. More interesting is the behavior of 
solutions where gravity is effectively coupled to the matter fields. In first
place, as it happens in asymptotically flat space, a critical value for the
  Newton constant   exists above which no regular 
monopole or dyon solution can be found. This effect was explained 
in asymptotically flat spaces \cite{LNW}-\cite{VG} by noting that as $\alpha_0$
 grows the mass of the monopole grows and its radius decreases 
 so that it finally becomes gravitationally unstable. Now, the presence of a
 cosmological constant enhances this effect and for this reason, 
 the critical value we find, $\alpha_0^c(\Lambda)$ is smaller than the
 asymptotically flat one.

\vspace{1 cm}

\noindent\underline{Acknowledgements}: This work is 
partially  supported by CICBA, CONICET (PIP 4330/96), ANPCYT
(PICT 97/2285).


\newpage
 
\begin{figure}
\centerline{ \psfig{figure=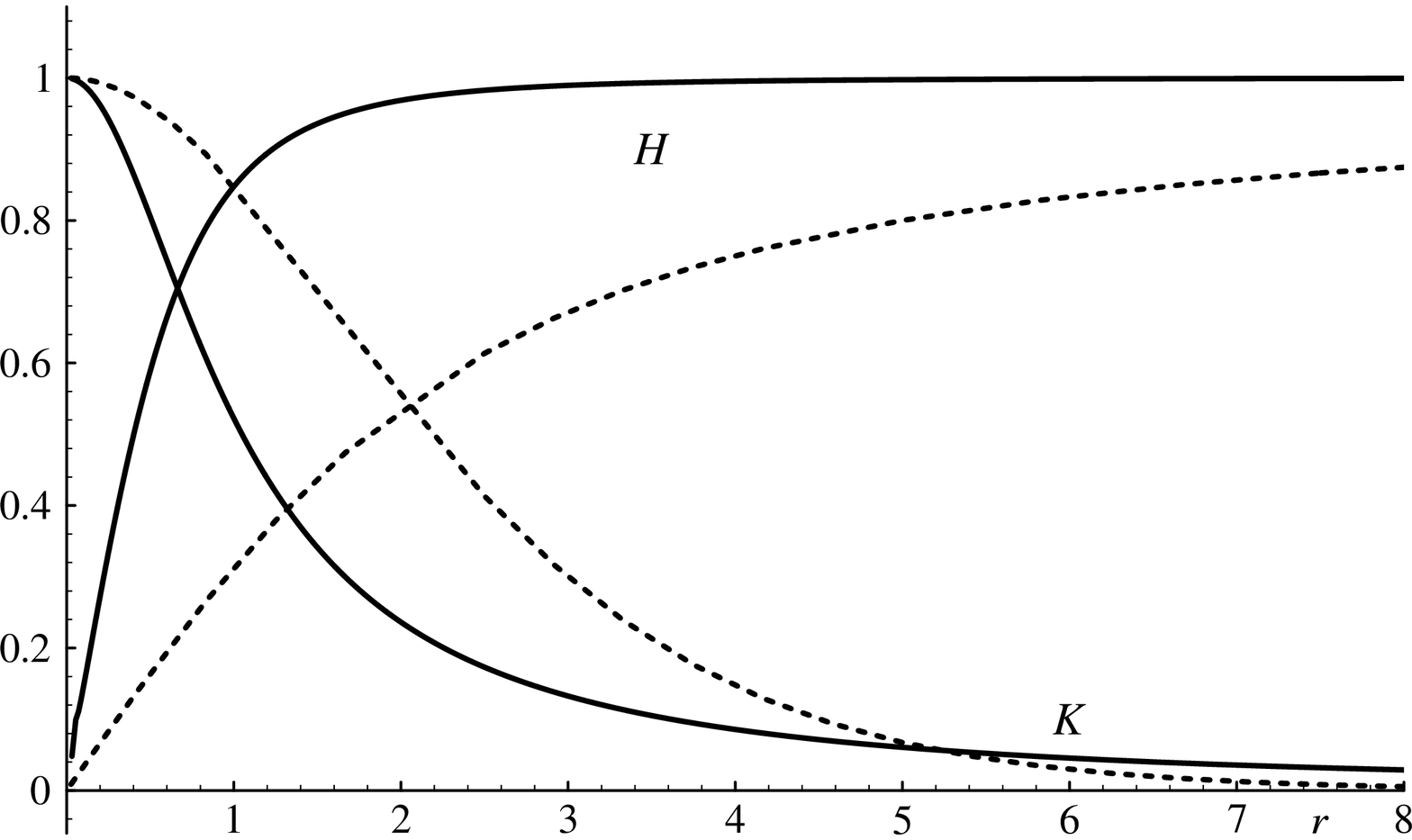,height=15cm,angle=0}}
\smallskip
\caption{ Plot of the functions $K(r)$ and $H(r)$ (in
dimensionless variables) for the monopole solution with
$\lambda=0$ and $\alpha_0 = 0$. 
The solid line corresponds to the solution with
$\gamma_0=1.0$ and the dashed line corresponds to  the 
BPS flat space solution.
 \label{fig1} }
\end{figure}

\newpage

\begin{figure}
\centerline{ \psfig{figure=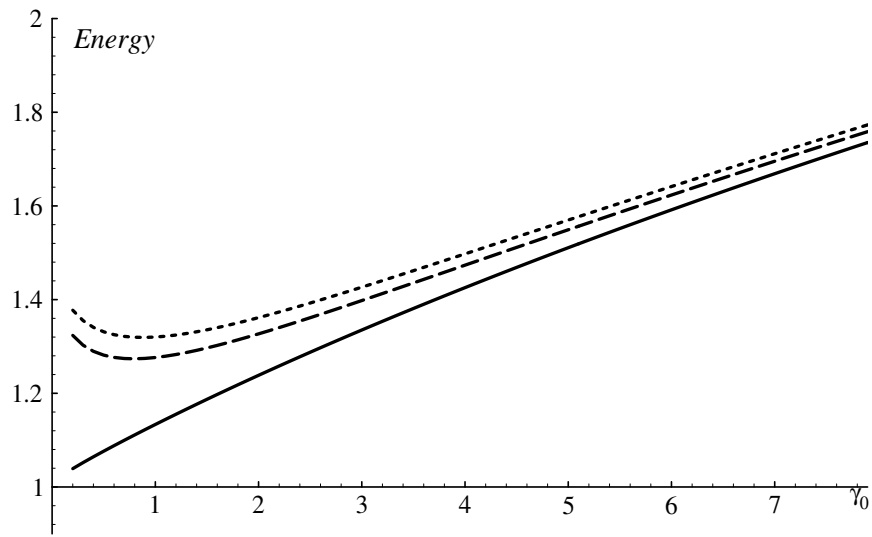,height=22cm,angle=0}}
\smallskip

\vspace{- 5 cm}
\caption{Energy of the monopole configuration as a function of
$\gamma_0$ for $h_0 = 1$ and different values of $\lambda$:
  $\lambda = 0$ (solid line), $\lambda = 10$ (dashed line)
and
$\lambda = 20$ (dotted line) in the $\alpha_0 = 0$ case.  
\label{fig-energy}}
\end{figure}

\newpage
\begin{figure}
\centerline{ \psfig{figure=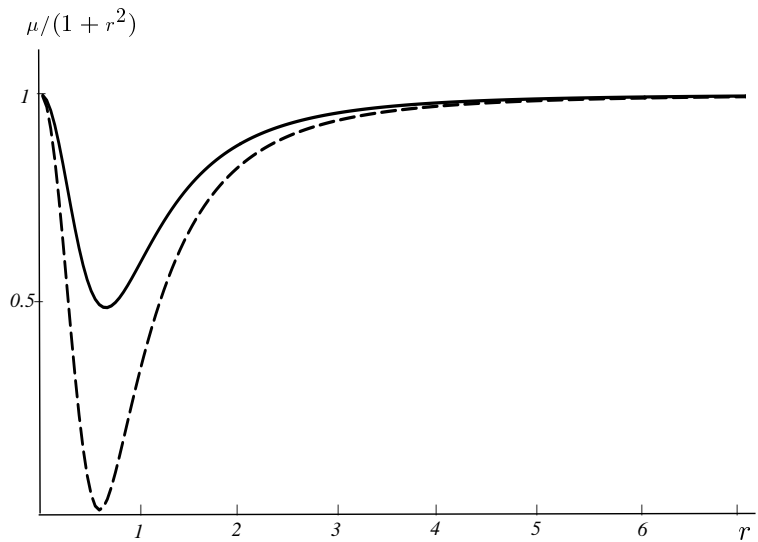,height=15cm,angle=0}}
\smallskip
\caption{ The solution for the
metric function ${\mu(r)}/{(1 +  r^2)}$ for fixed $\gamma_0 = 1$
and $\lambda = 0$. 
The solid line corresponds
to $\alpha_0 = 1$ and the dashed one to $\alpha_0 = \alpha_0^c 
\approx 1.371$.  
\label{fig-mu} }
\end{figure}
\newpage

\begin{figure}
\centerline{ \psfig{figure=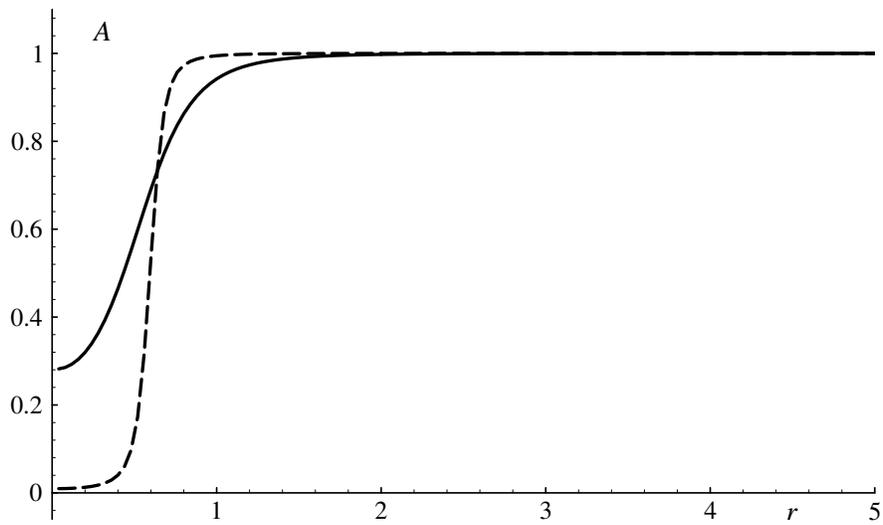,height=15cm,angle=0}}
\smallskip
\caption{ The solution for the
metric function $A(r)$ for fixed $\gamma_0 = 1$ and $\lambda=0$. The solid line corresponds
to $\alpha_0 = 1$ and the dashed one to $\alpha_0 = \alpha_0^c$.  
\label{A} }
\end{figure}


\end{document}